\documentstyle[12pt,aasms4]{article}
\catcode`\@=11
\def\gsim{\ifmmode{\mathrel{\mathpalette\@versim>}}
    \else{$\mathrel{\mathpalette\@versim>}$}\fi}
\def\lsim{\ifmmode{\mathrel{\mathpalette\@versim<}}
    \else{$\mathrel{\mathpalette\@versim<}$}\fi}
\def\@versim#1#2{\lower 2.9truept \vbox{\baselineskip 0pt \lineskip 
    0.5truept \ialign{$\m@th#1\hfil##\hfil$\crcr#2\crcr\sim\crcr}}}
\catcode`\@=12
\def\boxit#1{\vbox{\hrule\hbox{\vrule\kern3pt\vbox{\kern3pt#1\kern3pt}\kern3pt
             \vrule}\hrule}}

\def\yr{hbox{\rm yr}}

\def\ref{\par\noindent\hangindent=1truecm}
\def\feh{\hbox{{\rm [Fe/H]}}}

\def\hz{\hbox{$H_{\!\circ}$}}
\def\oz{\hbox{$\Omega_{\circ}$}}
\def\tgc{\hbox{$t_{\rm GC}$}}
\def\tgf{\hbox{$t_{\rm GF}$}}
\def\tz{\hbox{$t_{\circ}$}}

\def\mvto{\hbox{$M_{\rm V}^{\rm TO}$}}
\def\vto{\hbox{$V^{\rm TO}$}}

\def\mwd{\hbox{$M_{\rm WD}$}}

\def\3/2{\hbox{${3\over 2}$}}

\def\msun{\hbox{$M_\odot$}}

\def\yr-1{\hbox{${\rm yr}^{-1}$}}
\def\etal{\hbox{\it et al.\/$\,\,$}}

\def\mh1{\hbox{$M_{\rm H}^1$}}

\def\log{\hbox{\rm Log}}

\lefthead{Renzini et al.}
\righthead{White Dwarf Distance to NGC 6752}

\begin{document}

\title{THE WHITE DWARF DISTANCE TO THE \\ GLOBULAR CLUSTER NGC 6752
	(AND ITS AGE) \\ WITH THE HUBBLE SPACE 
         TELESCOPE$^*$}

\author{Alvio Renzini\altaffilmark{1,2}, Angela
Bragaglia\altaffilmark{3}, Francesco R. Ferraro\altaffilmark{3}, Roberto
Gilmozzi\altaffilmark{1}, Sergio Ortolani\altaffilmark{4}, J.B. 
Holberg\altaffilmark{5}, James Liebert\altaffilmark{6}, F.
Wesemael\altaffilmark{7}, \& Ralph C. Bohlin\altaffilmark{8}
}

\altaffiltext{1}{European Southern Observatory, D-85748 Garching b.
M\"unchen, Germany}
\altaffiltext{2}{Universit\`a di Bologna, Dipartimento di Astronomia,
I-40126 Bologna, Italy}
\altaffiltext{3}{Osservatorio Astronomico, I-40126 Bologna, Italy}
\altaffiltext{4}{Universit\`a di Padova, Dipartimento di Astronomia,
I-35122 Padova, Italy}
\altaffiltext{5}{University of Arizona, Lunar and Planetary
Laboratory, Tucson AZ 85721, USA}
\altaffiltext{6}{University of Arizona, Steward Observatory, Tucson AZ
85721, USA}
\altaffiltext{7}{Universit\'e de Montr\'eal, D\'epartement de Physique,
Montr\'eal, Qu\'ebec, Canada H3C 3J7}
\altaffiltext{8}{Space Telescope Science Institute, Baltimore, MD
21218, USA}

\vfill
$^*$ {Based on observations made with the NASA/ESA
{\it Hubble Space Telescope} (HST).}

\begin{abstract}
Deep {\it Hubble Space Telescope} (HST) observations with WFPC2 of the
nearby globular
cluster NGC 6752 have allowed us to obtain accurate photometry for the
cluster white dwarfs (WD). A sample of local WDs of known trigonometric
parallax and mass close to that of the cluster WDs have also been
observed with WFPC2. Matching the cluster and the local WD sequences
provides a direct  measure of the distance to the cluster: 
$(m-M)_\circ=13.05$, with an uncertainty less than $\pm0.1$ mag which
allows a substantial reduction in  the uncertainty in the  age of
the cluster. Indeed,
coupling this value of the cluster distance to the cluster metallicity,
helium abundance and $\alpha$-element enhancement [$\alpha$/Fe]=0.5
yields an age of 
15.5 Gyr and 14.5 Gyr using evolutionary models that do not include 
or do include helium diffusion, respectively. The
uncertainty affecting these age determinations is  $\sim 10\%$.
The majority of the cluster WDs appear to be of the DA variety, while
the color-magnitude location of two WDs is consistent with 
the DB type. This suggests a cluster DB/DA ratio  similar
to that of WDs in the solar neighborhood.

\end{abstract}


\section{Introduction}
The age of the universe $\tz$ is the obvious partner of the Hubble constant 
$\hz$. Together they set a constraint on $\oz$ if we believe the cosmological
constant $\Lambda$ to be zero, or on a combination of $\oz$ and $\Lambda$
if one is willing to accept  $\Lambda\ne 0$ cosmologies.
By general consensus, globular cluster ages provide  potentially the most 
accurate estimate of $\tz=\tgf+\tgc$, $\tgf$ being the
age of the universe when the Galaxy formed and $\tgc$  the present age of
galactic globular clusters. Since presumably $\tgf\simeq 1-2\,{\rm Gyr}\ll\tz$,
then $\tz\approx\tgc$, and  $\tgc$  provides a strict lower bound to $\tz$.
The age of Galactic globular clusters can be most accurately estimated by 
using the theoretical 
relation between age and the luminosity of the main sequence turnoff (TO), 
other methods being undermined by uncontrollable systematic errors (Renzini
1991, 1993).
For example, one can use a relation that fits the isochrones of VandenBerg 
\& Bell (1985):
$$\log t_9\simeq -0.41 + 0.37\,\mvto - 0.43\, Y - 0.13\,\feh,\eqno(1)$$
where $t_9$ is the age in Gyr units, 
$Y$ the helium abundance, $\feh$ the iron abundance in standard notations, and
 $\mvto$ the TO absolute visual
magnitude. In turn,  $\mvto = V^{\rm TO} -$ 
mod, where $V^{\rm TO}$ (the TO apparent 
magnitude) is the directly {\it observable} quantity, and mod is the cluster 
distance modulus.
This relation allows one to estimate the relative importance 
of the uncertainty in each of the four input quantities ($V^{\rm TO}$, mod,
$Y$, and [Fe/H]) in establishing the 
final uncertainty in the age determination. The current 
distances are typically affected by a $\sim$1/4
magnitude error in the modulus -- $\sigma($mod$)\simeq 0.^{\rm m}25$ --
which immediately translates into a $\sim 22\%$ error in
the derived cluster age ($\sim 3$ Gyr for an age of 15 Gyr). All other input
quantities convey substantially smaller errors.
The high photometric accuracy of CCDs now allows one to determine a cluster's
$V^{\rm TO}$ with an accuracy better than $0^{\rm m}.1$, which
translates into a $\sim 9\%$ error in age. The helium abundance 
is very well known, from either the R method, primordial nucleosynthesis, or
empirical determinations of the {\it pregalactic} abundance, which all
indicate $Y=0.23-0.24$ (e.g., Boesgaard \& Steigman 1985), and even a 
$\pm0.02$ 
uncertainty in $Y$ gives a 
negligible 2\% error in age. The metal content of the best studied
clusters is uncertain by $\sim$0.3 dex (most of it being systematic), which 
translates into a $\sim 9\%$ uncertainty in age. There is a problem with 
the {\it composition} of metallicity (e.g. enhanced [O/Fe], or [$\alpha$/Fe]), 
a point to which we shall return in Section 4.
Clearly the first concern is the error in the distance of the clusters,
and it is therefore instructive to recognize that distance determinations
dominate the error budget not just of the {\it kinematical} age of the 
universe (via $\hz$) but also of globular cluster ages. For the comparison 
of the two ages to be  unambiguous, the error in each of them 
must be  reduced as much as possible.
The {\it Hubble Space Telescope} (HST) {\it Key Project} is aimed at
achieving  $\sim 10\%$ accuracy on $\hz$ (Kennicut, Freedman, \& Mould 1995). 
We report  here  our own attempt at using HST observations to 
achieve similar accuracy on $\tgc$.

Using ground based observations, the distance to globular
clusters has been  estimated with either the RR Lyrae or
the subdwarf methods. Their limitations are extensively discussed by
e.g., Sandage \& Cacciari (1990) and Renzini (1991, 1993). Suffice it to 
mention here
that both methods are semi-empirical in nature, relying heavily on
theoretical models (e.g., pulsational, atmosphere, and stellar models),
and both require the metallicity of the calibrating
stars and of the clusters to be measured. Hence, the resulting estimate of
the distance  is  affected by both systematic errors
that are difficult to quantify, and by errors in metallicity which
can dominate the  age error budget.
In this paper we present the first attempt at determining the distance
to a globular cluster by using the white dwarf (WD) method (Renzini
1991 and ref.s therein),
that is essentially free from these limitations.

\section{The White Dwarf Method for Globular  Cluster Distances}

The basic idea of using WDs  as standard candles is very 
simple: to fit the WD cooling sequence of a globular cluster to an
appropriate empirical cooling sequence constructed using local WDs
with well determined trigonometric parallaxes.
The procedure is analogous to
the classical main sequence fitting to the local subdwarfs (e.g.,
Sandage 1970), but with some 
non-trivial advantages: the method does not involve metallicity 
determinations which inevitably come with their uncertainties, and there 
are no complications with convection.
In fact, WDs have virtually metal free
atmospheres, coming either in the DA or non-DA varieties (nearly pure hydrogen 
or pure helium, respectively).
Moreover, WDs are locally much more abundant than 
subdwarfs, and therefore accurate trigonometric parallaxes can be
obtained for a potentially much larger sample of calibrators. 
However, cluster WDs are very faint, with $V\gsim 24$ even in the
closest globular
clusters (De Marchi, Paresce, \& Romaniello 1995; Richer et al. 1995;
Cool, Piotto, \& King 1996). HST is therefore required to detect them
and to obtain photometric data of adequate accuracy. 

However, the location of the WD cooling sequence is  sensitive to the WD
mass. The mass of currently forming WDs in globular clusters should
therefore be estimated, and the local calibrating WDs must be chosen
among those matching cluster WDs. Theoretical WD models (e.g., Wood 
1995) give $\delta($mag$)\simeq 2.4\delta\mwd$ for the mass
dependence of WD magnitude at any given temperature (or color), and
therefore WD masses need to be determined with high accuracy for the 
method to provide competitive distances. 
On the one hand, the cluster $\mwd$
is very effectively constrained by four independent
 observations, namely: the luminosities of 1) the red giant branch tip,
 2) the horizontal branch, 3) the AGB termination, and 4) the post-AGB stars,
 which are all very sensitive to the mass of the hydrogen exhausted
core, and which
 consistently indicate $0.51\lsim\mwd\lsim 0.55\msun$, or
$\mwd=0.53\pm 0.02\;\msun$, virtually independent of metallicity (Renzini \&
 Fusi Pecci 1988). Therefore, also the WD method makes some use of
theoretical
models, but the quantities involved are the least model-dependent,
 essentially, the core mass-luminosity
 relation. All in all, the cluster $\mwd$ is perhaps the most robust
prediction of stellar
evolution theory applied to globular cluster stars.
In practice, the $0.02\msun$ uncertainty in the cluster $\mwd$ implies
an
uncertainty in the distance modulus of only $\sim 0^{\rm m}.05$, or a
5\%
uncertainty in age, which determines the superiority of WD method.

Local WDs are characterized by  a very narrow mass distribution
(1-$\sigma\simeq 0.1\msun$), with $<\!\mwd\! >=0.59\,\msun$ (Bergeron
 Saffer \& Liebert 1992; Bragaglia, Renzini, \& Bergeron 1995),
yielding a cooling sequence on the color-magnitude diagram having an 
intrinsically low dispersion (the cluster WD cooling sequence is
expected to be even narrower,  given the virtually identical 
masses of the current progenitors). For this project the local
calibrating WDs have been chosen according to the following criteria:
1) an accurate parallax and an accurate spectroscopic mass being
available, as close as possible to the cluster WD's mass 
and 2) $10,000\lsim T_{\rm eff}\lsim 20,000$ K, so as to match the
temperature range of the cluster WDs expected to be detected with our
HST observations. Table 1 lists the local WDs that have been used
in the present experiment.

\placetable{1}

\section{The HST Observations and Data Analysis}

For this experiment the cluster NGC 6752 was selected as being the
closest of the low-reddening clusters.
A field about 2 arcmin SE from the center of NGC 6752 was observed
with WFPC2,
through the F336W, F439W, F555W, and F814W filters.
Preliminary reductions have been performed on all the data, but we use 
here data only for the WDs in  
the less crowded of the four chips (WF4), and only for the two best 
exposed bands, F439W and F555W, with total exposure time 10,000s and
6,000s, respectively. The complete analysis will
be published elsewhere (Bragaglia et al. 1996).
Each exposure was processed through the standard HST--WFPC2 pipeline, 
including bias subtraction, dark correction and flat-fielding.
All images taken with the same filter have been aligned
and averaged using a MIDAS standard task to remove cosmic ray events. 
We then used the final, averaged F439W image 
to  inventory automatically  all the stars
$\sim 5\sigma$ above background.
The stellar
positions determined on this frame were then used
as input centers for the point spread function (PSF) fitting procedure
for the averaged F555W image.

Preliminary photometry of individual stars was performed on the averaged
F439W and F555W frames using ROMAFOT (Buonanno et al. 1983) 
in a version
specifically developed for handling HST data. In particular, the HST point
spread function is modelled by a Moffat function plus
a numerical map of residuals. The PSF parameters
have been determined analyzing the brightest uncrowded stars in each field.
On this preliminary color-magnitude diagram (including $\sim 1500$ objects)
we made a first selection in color, choosing all objects bluer than the
main sequence. We then eliminated obvious mistakes from this sample 
($e.g.$, remaining cosmic rays, blends, etc.), narrowing it to about
40 objects forming a fairly narrow sequence 
at the position expected to be populated by the
cluster WDs. In this way the candidate WDs have been  singled out for
further  study in order to achieve the best photometric accuracy.

To this end, the two-dimensional fitting was performed separately on each 
individual frame, obtaining up to 5 independent 
measurements.
Each  candidate WD has been examined by eye, and those 
contaminated by  diffraction spikes or light from
bright nearby stars have been rejected.
Only the best 21 objects have been retained and, for them, each
possibly compromised measure (from cosmic ray hits) has been excluded
for the average.

Aperture corrections to instrumental magnitudes have been applied
using an aperture radius of $0''.5$, as suggested by Holtzman et al. (1995).
For each filter, about a dozen of the brightest, unsaturated
and  isolated stars were examined on one single frame;
the differences between the $0''.5$ aperture magnitudes and the fitting
magnitudes for these reference stars were averaged, and the resulting
aperture correction applied to all WD candidates in the WF4 chip.
The final instrumental magnitudes, obtained by averaging the
independent measures in each filter, have been reduced to 1 second 
exposures. The resulting CMD for the
21 WDs  is presented in Fig. 1a that also shows
the individual photometric errors for each WD candidate.
The errors have been computed  as the root mean square of the frame to frame
scatter of the instrumental magnitudes of each star.
The photometric errors in each
filter have then been added in quadrature  to produce the error in color.
Reddening corrections have been applied to the cluster WDs, adopting $E(B-V)=
0.04\pm0.02$ (Penny \& Dickens 1986). This corresponds to
E(F436W-F555W)= 0.036 and A(F555W)= 0.13 (Holtzman et al. 1995).

\placefigure{fig 1}

The local calibrating WDs  have also been observed
with WFPC2, exposing each of them in each of the four CCD chips
through the same four filters as the cluster, thus totalling 16, 
S/N$\simeq 100$ WFPC2 observations per star, though only WF4 data have
been used here.
On these frames aperture photometry of each WD image was obtained
using a $0''.5$ aperture radius. The resulting magnitudes have
also been reduced to 1 second, so that cluster and field WDs magnitudes
are completely homogeneous. The absolute instrumental magnitudes of these
WDs have then been obtained from their trig parallaxes
(Table 1), and their location in the absolute color-magnitude
diagram is displayed in Fig. 1b.

\section{The WD Cooling Sequence of NGC 6752, its Distance and Age}

The WD cooling sequence in Fig. 1a appears as 
a straight line in the diagram. Four stars lie definitely
outside this main WD sequence.
The two stars above the sequence
might overlap with a lower main sequence star in either a
physical or a projection binary; this is confirmed by their
separation from the main WD sequence being much larger in
diagrams involving the $I$ band photometry (F814W).
A comparison with
panel (b) suggests that the two stars lying below the main WD 
sequence belong to the DB variety, and this is further reinforced by
their behavior in all diagrams involving also the F336W and F814W
colors (Bragaglia et al. 1996). Therefore, we  identify
the main WD sequence with the sequence of WDs of the DA
variety, and notice that the DB/DA ratio of the cluster (roughly $\sim 10\%$)
appears to be consistent with that of the WD population in the solar 
neighborhood (e.g., Sion 1984).
Having excluded these four outliers, we then proceed to obtain
the distance of the cluster.

A vertical shift by $\delta$(F555W)=-13.05 brings the the cluster WD sequence
to overlap the local calibrating WDs of the DA variety, 
 as displayed in Fig. 1c, and we conclude that the distance
modulus of NGC 6752 is 
$(m-M)_\circ = 13.05$ (cf. mod=13.12 as reported by  Djorgovski, 1993). 
The formal uncertainty of the fit is very low ($\sim 0^{\rm m}.025$).
When taking into account uncertainties in
the relative cluster and local WD photometry, reddening, parallax and
average mass offset of the calibrating WDs we conservatively estimate
the overall uncertainty in the distance modulus to be less than $\pm 0.1$ mag.
It is worth emphasizing that this determination of the distance
modulus does not require absolute
photometric calibrations, and for this reason we prefer to stick to the
instrumental magnitude scale.
With this measure of the distance modulus we  proceed to determine
the absolute Johnson $V$ magnitude of the  main sequence
turnoff, and then the cluster age. For the cluster parameters we
adopt:
$\vto=17.4\pm 0.07$ and $A_{\rm V}=0.12\pm 0.06$ (Penny \& Dickens
1986), [Fe/H]=--1.54$\pm 0.3$ (Zinn 1985), and
$Y=0.23\pm 0.02$ (Boesgaard \& Steigman 1985). 
Thus, $\mvto=4.23$, and entering equation
(1) one gets a cluster age of 18.0 Gyr. This assumes solar proportions
for the cluster heavy elements. Under the same assumption, 
the models of Salaris et al. (1993)
[see their equation (6)] yield an age of 17.8
Gyr. However, there is now ample evidence that the $\alpha$-elements
(i.e., O, Ne, Mg, Si) are enhanced relative to iron in metal poor
halo stars and clusters, with [$\alpha/$Fe]=0.4--0.6 at the
metallicity of NGC 6752 (e.g., Bessell, Sutherland, \& Ruan 1991).

Salaris et al. have shown  that what matters in the
age-$\mvto$ relation is the overall heavy element abundance [M/H], rather
than the detailed distribution, with
[M/H]=[Fe/H]+log(0.638$f_\alpha$+0.362),
where $f_\alpha$=dex[$\alpha$/Fe]. Thus, for [$\alpha$/Fe]=0.4 and 0.6
the models of Salaris et al. give a cluster age of 16.1 and 15.3 Gyr, 
respectively. For [$\alpha$/Fe]=0.6 (i.e., [M/H]=--1.1) we estimate an
age of 15.2 Gyr from Fig. 7 in Bergbush \& VandenBerg (1992), once
more showing the good agreement between different sets of stellar models
when identical cluster parameters and assumptions are adopted.
This latter estimate assumes no helium diffusion inside stars during
their main sequence lifetime. When helium diffusion is allowed,
Bergbush \& VandenBerg models yield an age of 14.0 Gyr for [M/H]=--1.1.
The error to attach to these age determinations can be  estimated from
the propagation of the errors in the various input paramenters 
$V^{\rm TO}_{\circ}$, 
$(m-M)_\circ$, [M/H], and $Y$. We estimate the overall error to be
$\sim 10\%$ (a detailed error analysis will be presented along with the
full data set in Bragaglia et al. 1996), with a systematic uncertainty
related  to helium diffusion of $\sim\pm 0.5$ Gyr. In summary, adopting the
central value [$\alpha$/Fe]=0.5, one obtains an age of $15\pm1.5\pm0.5$ Gyr
(random plus systematic uncertainties).

Unlike the quest for $\hz$, different groups have always estimated
globular cluster ages that are in substantial agreement with each other.
Not surprisingly, 
the cluster age we have derived is in tight
agreement with all other recent estimates (e.g. Bolte \& Hogan 1995,
and references therein). Yet, we claim to have achieved a sizable
decrease
of the error affecting this determination, from $\sim 25\%$ to below
$\sim 10\%$. When the age of the universe at the epoch of the
formation of NGC 6752 is taken into account (1--2 Gyr), we end up with
a present age of the universe of that can hardly be lower than 15 Gyr.
When coupled with current
estimates of the Hubble constant the
well known age problem is encountered, with high-$\oz$ cosmological
models being disfavored, and the cosmological constant $\Lambda$
making an entrance many would prefer not witnessing. We restrain from
embellishing further on this issue, and rather focus on what can still
be done to put globular cluster ages on even firmer grounds.

Further improvements in all steps involved in the age determination
may include tightening down further the
accuracy of cluster photometry and stellar abundances. 
Improved trig parallaxes of nearby WDs and a wider
number of such calibrators to be observed with WFPC2 would also be of
interest. All this together may somewhat reduce the {\it random} error
below $\sim 10\%$.
Yet, the main surviving uncertainty 
is perhaps of rather systematic nature. After all the use of stellar
models is unavoidable in the age dating process, and they still
need to be thoroughly tested before the resulting globular cluster
ages can be regarded as definitively established.
However, in spite of the pressure on this issue
no  obvious shortcoming of stellar models
has yet emerged, which is able to significantly affect the derived globular
cluster ages. The most crucial test to be adequately performed is
perhaps one in which theoretical and empirical luminosity functions
are compared, especially for the luminosity range going from the
turnoff to the lower
red giant branch (e.g., Renzini \& Fusi Pecci 1988; Renzini 1991). Any
effect able to accelerate central hydrogen exhaustion and/or an early
expansion and cooling of the envelope (hence our stellar clock
readings) should leave its imprint in the 
luminosity function that would show up as an excess of stars in the
luminosity
range just above turnoff. Very extensive, complete, uncontaminated and
photometrically accurate samples of cluster stars are needed for this 
fundamental check. 

\acknowledgments

We are grateful to G. Piotto for extensive discussions on the
best strategies to obtain accurate stellar photometry from HST data,
to R.  Lucas, M. Mutchler, and A. Suchkov
at STScI for their invaluable help in getting our HST program properly
set up, and to the referee for helpful comments. This project was
supported in part by the Italian Space Agency (ASI), by 
NASA through grant 
GO-05439.01-93A from STScI (to JBH and
JL),  and  by
NSERC Canada and FCAR Qu\'ebec (to FW).

%
%

\clearpage
\begin{table*}
\begin{center}
\caption{\bf The Local Calibrating White Dwarfs}
\begin{tabular}{cccccccc} 
WD & $\pi$ &ref.  & M/M$_{\odot}$ &ref.\\
\tableline
DA WDs: &&&&&&&\\
0839$-$327 & $0.1123\pm 0.0072$ & 1 &$0.553\pm 0.063$ & 3\\ 
1935+276   & $0.0561\pm 0.0029$ & 1 &$0.512\pm 0.013$ & 5\\ 
1327$-$083 & $0.0611\pm 0.0028$ & 1 &$0.502\pm 0.017$ & 3\\ 
2341+322   & $0.0559\pm 0.0017$ & 1 &$0.494\pm 0.021$ & 5\\ 
2126+734   & $0.0433\pm 0.0035$ & 2 &$0.513\pm 0.012$ & 5\\ 
DB WDs: &&&&&&&\\
0002+729   & $0.0291\pm 0.0047$ & 1 &$0.60\pm 0.03$ & 6\\  
1917$-$077 & $0.1010\pm 0.0026$ & 1 &$0.55\pm 0.05$ & 7\\
\tableline
\end{tabular}
\tablenotetext{}{References: 1: Van Altena et al. (1991); 2: Harrington
\& Dahn, (1980); 3: Bragaglia et al. (1995); 4: Bergeron et al. (1995);
5: Bragaglia \& Bergeron (1996); 6: Beauchamp (1995); 7: Oswalt et al.
(1991).}
\end{center}
\end{table*}


\clearpage


\figcaption{(a) The instrumental color-magnitude diagram for the
cluster white dwarfs detected on the CCD chip \#4 of the WFPC2; (b):
the instrumental {\it absolute} color-magnitude diagram for the local,
calibrating white dwarfs (from WF4 data only) of known trig parallax
that are
listed in Table 1 (in order of decreasing F555W luminosity to
allow cross identification).
WDs of the DA and DB varieties are represented by different symbols;
(c): the instrumental color-magnitude diagram of the cluster
and local WDs, with the former ones having been shifted in magnitude
to match the local sequence. This operation delivers the distance
modulus of the cluster: $(m-M)_\circ =13.05$. The straight line is a linear
fit to the cluster WD sequence.\label{fig1}}
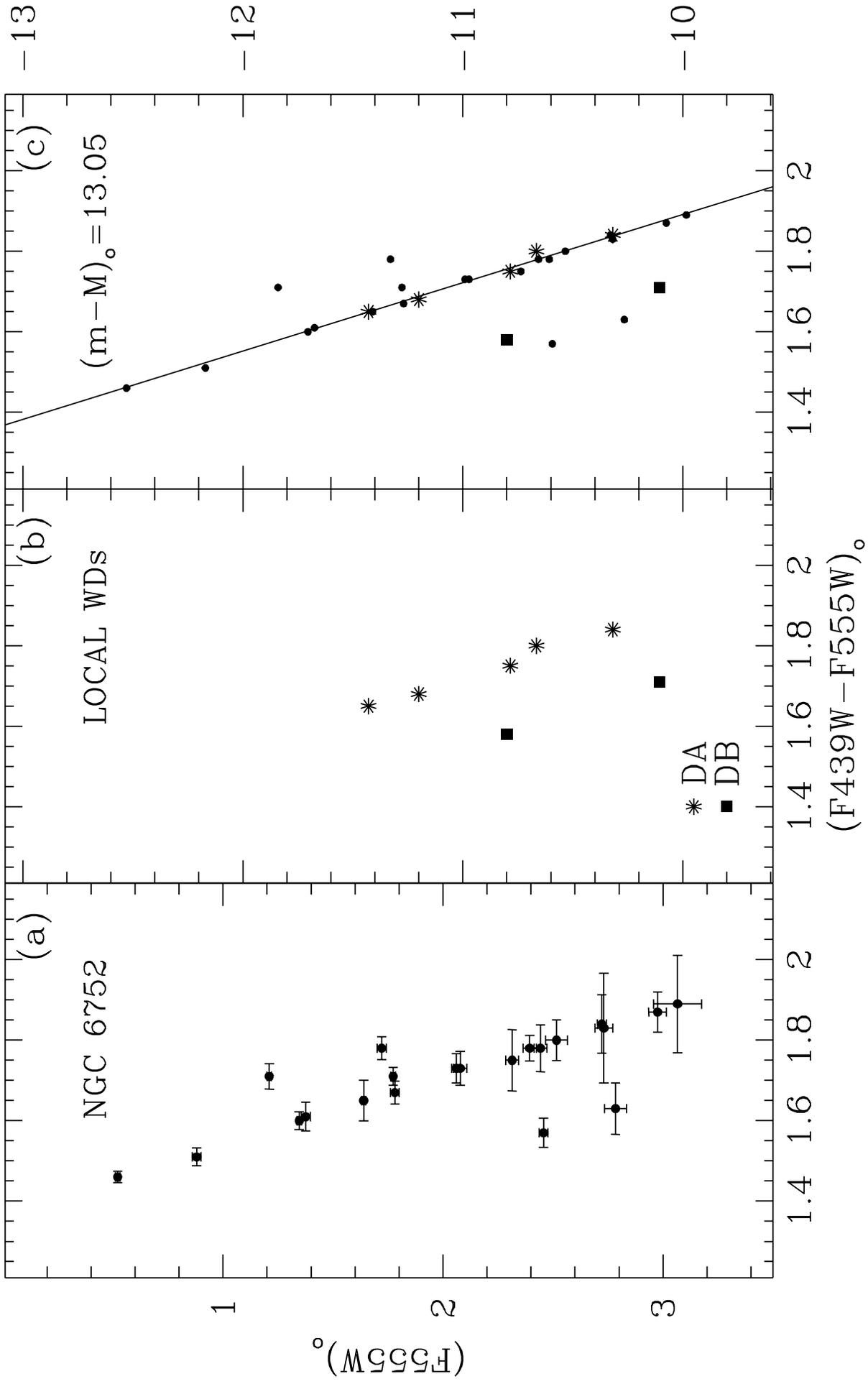
\end{document}